\documentstyle[a4,11pt,psfig]{article}
\title{Complementary approaches to the  \\ 
{\em ab initio} calculation of melting properties}
\author{D. Alf\`{e}$^{\star \,\dagger}$, M. J. Gillan$^\dagger$ and
G. D. Price$^\star$ \medskip \\ $^\star$Geological Sciences Department,
University College London \\ Gower Street, London WC1E 6BT, UK \medskip \\
$^\dagger$Physics and Astronomy Department, University College London
\\ Gower Street, London WC1E 6BT, UK}

\begin{document}
\maketitle
\section*{Abstract}
Several research groups have recently reported {\em ab initio}
calculations of the melting properties of metals based on density 
functional theory, but there have been unexpectedly large disagreements
between results obtained by different approaches. We analyze
the relations between the two main approaches, based on calculation
of the free energies of solid and liquid and on direct simulation
of the two coexisting phases. Although both approaches rely
on the use of classical reference systems consisting of
parameterized empirical interaction models, we point out that
in the free energy approach the final results are independent
of the reference system, whereas in the current form of the
coexistence approach they depend on it. We present a scheme
for correcting the predictions of the coexistence approach
for differences between the reference and {\em ab initio}
systems. To illustrate the practical operation of the scheme,
we present calculations of the high-pressure melting properties of 
iron using the corrected coexistence approach, which agree closely with
earlier results from the free energy approach. A quantitative
assessment is also given of finite-size errors, which we show
can be reduced to a negligible size.


\section{Introduction}
\label{sec:intro}

The study of solid-liquid equilibrium by computer simulation has a
long history, going back to the classic work of Alder and
Wainwright~\cite{alder57} on the hard-sphere system in the 1950's.
Many techniques have been used to determine the pressure-temperature
relation at equilibrium and other melting properties, such as the
volume and entropy of fusion. In the last few years, there has been an
upsurge of interest in the accurate {\em ab initio} treatment of the
melting properties of real
materials~\cite{sugino95,dewijs98,alfe99,alfe02,vocadlo02,laio00,belonoshko00},
which has focused attention again on the techniques used to locate the
melting transition. Recent {\em ab initio} work has been based on two
main approaches. The first locates the melting transition by requiring
equality of the Gibbs free energies, which are calculated {\em ab
initio} for liquid and
solid~\cite{sugino95,dewijs98,alfe99,alfe02,vocadlo02}; we call this
the `free energy' approach. The second proceeds by fitting a potential
model to {\em ab initio} calculations and using this to simulate a
system containing liquid and solid in
coexistence~\cite{laio00,belonoshko00,morris94}; we call this the `coexistence'
approach. If appropriate measures are taken, the two approaches should
clearly give the same results. The purpose of this paper is to analyze
the relation between the two approaches and to propose what these
`appropriate measures' should be. We illustrate our analysis by
presenting {\em ab initio} calculations on the high-pressure melting
of iron performed using the coexistence approach, which we compare
with earlier free-energy results from the same {\em ab initio}
technique~\cite{alfe99,alfe02}.  We shall show that, once the
necessary corrections have been applied, the two approaches yield the
same results.

The free-energy approach has been well established for many years for
calculations based on classical interaction 
models~(see e.g. Refs.~\cite{broughton87,foiles89,mei92}). Typically
the procedure for the solid has been to start from the
harmonic approximation at low temperatures; the free
energy of the high-temperature anharmonic system is then
obtained by using the Gibbs-Helmholtz relation to integrate
the internal energy given by molecular dynamics (MD) 
simulation~\cite{broughton87,foiles89}.
Alternatively, the anharmonic free energy has sometimes
been obtained by thermodynamic integration starting from
a reference model such as the Einstein solid~\cite{mei92}. For the liquid,
a common procedure has been to obtain the free energy at one
thermodynamic state from the work done in 
reversible expansion~\cite{broughton87,mei92}
to low density or heating to high temperature. The Gibbs-Helmholtz
relation is then used to obtain the free energy at other states.
We also mention an important alternative approach, known as
`Gibbs-Duhem' integration, which allows the boundary between
coexisting phases to be directly mapped out~\cite{agrawal95,sturgeon00}.

In early work, the interaction models were parameterized by fitting to
experimental data, but recent years have seen a major shift towards
the calculation of melting
properties~\cite{sugino95,dewijs98,alfe99,alfe02,vocadlo02,laio00,belonoshko00}
from {\em ab initio} methods based on density-functional theory
(DFT)~\cite{generaldft}.  In DFT, the total energy function of a system
is determined by the approximation used for exchange-correlation
energy $E_{\rm xc}$. An important ambition then follows: the
determination of melting properties with no statistical-mechanical or
other approximations except those inherent in $E_{\rm xc}$ itself.
This raises major new issues, because it is extremely costly to
perform {\em ab initio} MD (AIMD) simulations on large systems long
enough to reduce statistical errors to an acceptable level. The cost
is particularly great for crystals, since extensive electronic
$k$-point sampling may be needed. Long accepted methods like
reversible expansion to low density are impracticable with
AIMD. Instead, it is essential to use empirical interaction models
that closely mimic the DFT system. The usual techniques are then
employed to treat these models, and in some {\em ab initio} work
thermodynamic integration is used to obtain the free energy difference
between the model and DFT
systems~\cite{sugino95,dewijs98,alfe99,alfe02,vocadlo02}. In parallel
with these developments, the advantages of avoiding intricate
free-energy calculations by using the coexistence approach have made
this route popular~\cite{laio00,belonoshko00,morris94}.  Again, the
simulations must be based on empirical models matched to DFT data.

The present work was stimulated by recent reports of {\em ab initio}
or nearly {\em ab initio} calculations on the melting of a number of
materials, including Si~\cite{sugino95}, Al~\cite{dewijs98,vocadlo02},
Cu~\cite{belonoshko00a} and
Fe~\cite{alfe99,alfe02,laio00,belonoshko00}. Our own work on
Al~\cite{dewijs98,vocadlo02} and Fe~\cite{alfe99,alfe02} was based on
the free-energy approach, which used an inverse-power reference model,
with thermodynamic integration to calculate the difference between the
{\em ab initio} and reference free energies. A major effort was made
to ensure that finite-size and other systematic errors were reduced to
an almost negligible level. The other {\em ab initio} work on Fe, by
two independent groups, used the coexistence
approach~\cite{laio00,belonoshko00}. The results for the melting curve
differed substantially from each other and from our results, and we
urgently need to understand the reasons for the disagreements. We aim
to shed light on possible reasons here.

We shall explore a number of technical issues. The first concerns the
correction of the coexistence method for errors due to the
difference between the {\em ab initio} system and the empirical
model that mimics it. A second issue is the fitting of models
to {\em ab initio} data. We have already discussed this
in depth for the free-energy approach, so here we focus mainly on
coexistence. We want to study what physical quantities should be
fitted, and how to tell if the fit is good enough. A third
important issue concerns finite-size errors, which arise
mainly from the limited system sizes that can be handled {\em ab
initio}. It has been claimed~\cite{belonoshko00a} that 
the coexistence approach suffers
less from size errors than the free-energy approach. We shall
demonstrate that in fact this type of error affects
both approaches in essentially the same way. One thing we
shall not do is to pass judgment on which method is `better', since
we shall argue that the answer depends on what one wishes to
achieve, and that both approaches are vital. However, we
shall comment on the strengths and weaknesses of both.


In the next Section, we define the problem, summarize briefly
the free-energy and coexistence approaches as they have
been applied in practical {\em ab initio} calculations,
analyze the corrections that need to be applied in the two
approaches, and discuss size errors. Section~\ref{sec:iron}
then presents our new calculations on the high-pressure
melting properties of Fe using the coexistence approach. We shall
compare with our earlier free-energy results and show the practical
necessity of the corrections outlined in Sec.~\ref{sec:theory}.
The final Section gives further discussion and summarizes our
conclusions.

\section{Theory of the two approaches}
\label{sec:theory}

\subsection{Definition of the problem}
\label{sec:definition}

We start by defining the problem to be addressed. At a given pressure
$p$, two phases are in thermodynamic equilibrium when their
Gibbs free energies $G ( p , T )$ are equal. We regard $G$ as derived
by the relation $G = F + p V$ from the Helmholtz
free energy $F ( V , T )$. The {\em ab initio} value of $F$
at any volume and temperature, denoted by $F_{\rm AI} ( V, T )$, is
given in classical statistical mechanics by:
\begin{equation}
F_{\rm AI} ( V , T ) = - k_{\rm B} T \ln \left\{ \frac{1}{N! \Lambda^{3 N}}
\int d {\bf r}_1 \ldots d {\bf r}_N \, 
e^{- \beta U_{\rm AI} ( {\bf r}_1 , \ldots {\bf r}_N )} \right\} \; ,
\end{equation}
where $U_{\rm AI} ( {\bf r}_1 , \ldots {\bf r}_N )$ is
the {\em ab initio} total energy as a function of the positions
${\bf r}_1 , \ldots {\bf r}_N$ of the $N$ nuclei, which range
over the volume $V$ of the cell containing the system. For the purpose
of treating phase equilibria, we need the free 
energy per atom $F_{\rm AI} / N$ in the
thermodynamic limit $N \rightarrow \infty$, $V \rightarrow \infty$ at
constant number density $N / V$.

Even though the coexistence approach does not work directly
with free energies, we shall show later that the errors that need
to be overcome in this approach, as well as in the free-energy
approach, can be formulated as free-energy errors. But the
errors in calculating $F_{\rm AI}$ are of two kinds: electronic-structure
errors, i.e. imperfections in the calculation of the {\em ab initio}
total energy $U_{\rm AI}$ at each set of nuclear positions; and
statistical-mechanical errors, i.e. errors in calculating $F_{\rm AI}$
from the given $U_{\rm AI}$, and in taking the thermodynamic limit. 

For the purposes of this work, electronic-structure errors
are irrelevant: our sole concern is the treatment of phase
equilibrium using some {\em given} algorithm for computing
$U_{\rm AI}$. The `free energy' and `coexistence'
routes  differ only in the way they address the statistical
mechanics. Our problem is therefore to assess and compare the
ways that the two approaches control the purely statistical-mechanical
errors. When we come to the practical calculations in Sec.~\ref{sec:iron},
we shall compare the results of the two approaches applied
using exactly the same algorithm for $U_{\rm AI}$.

\subsection{The free-energy approach}
\label{sec:free}

\begin{sloppypar}
In the free-energy approach~\cite{sugino95,alfe99,alfe02,alfe01}, 
we use a reference system
with total energy function $U_{\rm ref} ( {\bf r}_1 , \ldots {\bf r}_N )$,
whose solid and liquid free energies $F_{\rm ref}$ are calculated
for very large systems, so that finite-size errors are negligible.
Then the only demanding problem is the calculation of the
difference $\Delta F \equiv F_{\rm AI} - F_{\rm ref}$,
which is accomplished by thermodynamic integration~\cite{frenkel96}:
\begin{equation}
\Delta F = \int_0^1 \langle \Delta U \rangle_\lambda \,
d \lambda \; ,
\end{equation}
where $\Delta U \equiv U_{\rm AI} - U_{\rm ref}$, and
the thermal average $\langle \, \cdot \, \rangle_\lambda$
is taken in the ensemble generated by the total-energy
function $U_\lambda \equiv ( 1 - \lambda ) U_{\rm ref} +
\lambda U_{\rm AI}$. In practice, $\langle \Delta U \rangle_\lambda$
is computed as a time average in an AIMD
simulation whose dynamics is governed by $U_\lambda$. The main kinds
of error are: statistical errors in the evaluation of
$\langle \Delta U \rangle_\lambda$; integration
errors due to inadequate numbers of $\lambda$ points; and finite-size
errors. In the systems studied so far by this approach, all these
errors can be brought under tight control.
\end{sloppypar}

It is a crucial feature of this approach that $F_{\rm AI}$, calculated
as $F_{\rm ref} + \Delta F$, does not depend on the choice of
reference system, provided all technical tolerances are set so as to
suppress the errors we have just mentioned.  However, for reasons
expounded in detail elsewhere~\cite{alfe99,alfe02,vocadlo02}, the
choice of $U_{\rm ref}$ has a major influence on the overall
computational effort, which is minimized by reducing as much as
possible the strength of the fluctuations $\left\langle \delta \Delta
U^2 \right\rangle_\lambda / N$, where $\delta \Delta U \equiv \Delta U
- \langle \Delta U \rangle_\lambda$.  In particular, minimization of
this fluctuation strength is important in ensuring that $F_{\rm ref}$
accounts for almost all the free energy, so that size errors in the
small residue $\Delta F$ are negligible. This can be achieved by using
a parameterized reference model whose parameters are adjusted so as to
minimize $\langle \delta \Delta U^2 \rangle_\lambda / N$, as in our
work on the melting of Fe. In this approach, it is an advantage that
different reference systems can be used for solid and liquid, since
for many materials it may be difficult to create $U_{\rm ref}$
functions that mimic $U_{\rm AI}$ with high precision in both phases.

\subsection{The coexistence approach}
\label{sec:coexist}

The coexistence approach is also based on a reference model that
mimics the {\em ab initio} system. (This model is sometimes called by
names such as `optimised potential model'~\cite{laio00,ercolessi94},
but here we give it the same name as in the free-energy approach.)
Various ways have been used to fit the reference model to the {\em ab
initio} system. Since it will be relevant later, we note that recent
coexistence work on Fe has used the `force-matching' procedure of
Ercolessi and Adams~\cite{ercolessi94}, in which the reference
parameters are adjusted so that the {\em ab initio} atomic forces are
reproduced as well as possible for representative sets of atomic
positions.

There are also several ways of using the reference model to simulate
coexisting phases, and hence to determine the phase boundary
between them. In the work of Morris {\em et al.}~\cite{morris94},
coexisting solid and liquid Al
were simulated with the total number of atoms $N$, volume $V$
and internal energy $E$ fixed. They showed that, provided
$V$ and $E$ are appropriately chosen, the two phases
coexist stably over long periods of time, and the average pressure
$p$ and temperature $T$ in the system give a point on the
melting curve. The underlying concept is that the mean
volume per atom $\bar{v} \equiv V / N$ is given
by:
\begin{equation}
\bar{v} = ( 1 - x_l ) v_s ( p ) + x_l v_l ( p ) \; ,
\label{eqn:vol_per_atom}
\end{equation}
where $x_l$ is the fraction of the atoms in the liquid phase,
with $v_s ( p )$ and $v_l ( p )$ the volumes per atom in
the coexisting solid and liquid as a function of pressure. For
fixed $\bar{v}$, the pressure $p$  traverses a certain
range as $x_l$ goes from 0 to 1. As $x_l$ varies in this way, the mean
internal energy per atom $\bar{e} \equiv ( 1 - x_l ) e_s ( p ) +
x_l e_l ( p )$ also traverses some range
(here, $e_s ( p )$ and $e_l ( p )$ are the internal
energies per atom in the two phases). 
Provided $\bar{e}$ lies in this range for the given
$\bar{v}$, the simulation will yield stably coexisting solid
and liquid. An alternative procedure would be to simulate
at constant $(N, V, T)$. Then coexistence will be achieved
for a given $\bar{v}$ provided $T$ is chosen so that the
corresponding $p$ on the melting line curve falls in the range specified by
Eq.~(\ref{eqn:vol_per_atom}). Yet another approach was used in the work of
Laio {\em et al.}~\cite{laio00} on the high-pressure melting of Fe; 
this used constant-stress simulations, with enthalpy almost exactly
conserved. The approach of Belonoshko 
{\em et al.}~\cite{belonoshko00} is different
again. Here, the $(N, p, T)$ ensemble is used.
The system initially contains coexisting solid and liquid, but since
$p$ and $T$ generally do not lie on the melting line, the system
ultimately becomes entirely solid or liquid.  This approach does not
directly yield points on the melting curve, but instead provides upper
or lower bounds, so that a series of simulations is needed to locate
the transition point.
Whichever scheme is used, some way is needed of monitoring
which phases are present. In the $(N, V, E)$ method of Morris
{\em et al.}~\cite{morris94}, graphical inspection of particle
positions appears to have been used, supplemented by
calculating of radial distribution functions to confirm
the crystal structure of the solid. In the $(N, p, T)$ method
of Belonoshko {\em et al.}~\cite{belonoshko00}, the primary
diagnostic is the discontinuity of volume as the system transforms
from solid to liquid.

The coexistence calculations presented here on the high-pressure
melting of Fe employ the $(N, V, E)$ method applied to reference
systems consisting of the `embedded-atom model'~\cite{daw93} fitted to
{\em ab initio} data. For geophysical reasons, we are interested in
pressures near that at
the boundary between the Earth's inner and
outer cores, namely 330~GPa~\cite{poirier91}. In this 
pressure region, the most stable
crystal structure just below the melting curve is believed to be
hexagonal close packed (h.c.p.)~\cite{vocadlo99,vocadlo00}, and we
assume here that melting occurs from this phase.  We start with a
simulation cell containing only the h.c.p. solid, with the basal plane
parallel to one face of the cell, and the system is allowed to
thermalize at a temperature where coexistence is expected. At some
instant of time, the simulation is stopped, and the atoms in one half
of the system are held fixed; the boundary between the two halves is
taken parallel to the h.c.p. basal plane. The atoms in the other half
are raised to a very high temperature, and dynamical evolution of
these atoms is allowed to proceed so that melting occurs. With this
half of the system molten, its temperature is now reduced to the
original value, the atoms in the other half still being
fixed. Finally, the atoms in the fixed half are given thermal
velocities and released, and the whole system is allowed to evolve
freely. The system is monitored by calculation of the average number
density in slices of the cell taken parallel to the boundary between
solid and liquid. As we shall show, the density in the solid part is a
periodic function of slice number, while in the liquid it fluctuates
rather weakly about its average value.  The total energy, temperature
and pressure are, of course, monitored throughout the simulation. A
feature of the $(N, V, E)$ method is that at equilibrium the stress in
the solid phase will generally not be hydrostatic, and manual
adjustment of the cell parameters is needed to achieve hydrostatic
stress.  We shall show in Sec.~\ref{sec:iron} that this does not
present a problem.

\subsection{Correcting the coexistence approach}
\label{sec:corrections}

Both the free-energy and the coexistence approaches are subject
to errors, which need to be assessed and corrected for. It may be
too costly to perform all the {\em ab initio} calculations
with the precision needed to obtain accurate melting properties,
so that there are errors due to inadequate $k$-point sampling,
incompleteness of the basis set, or other approximations. 
In the coexistence approach, inevitable
differences between the reference and {\em ab initio} total energies
will create errors in the melting properties, which need to be
evaluated. The correction of errors in the free-energy approach
has been extensively discussed 
elsewhere~\cite{alfe02,alfe01}, so our main concern
here is correction of the coexistence approach.

Assuming the {\em ab initio} energy is calculated with adequate
precision, the main error comes from differences between $U_{\rm AI}$
and $U_{\rm ref}$.  The key question posed here is therefore: how are
predicted melting properties changed by small changes in the total
energy function? (System-size errors will be discussed separately
later.) Related questions have been discussed
before (see e.g. Ref.~\cite{sturgeon00}), so we give 
only a brief summary of the main points.

The difference $U_{\rm AI} - U_{\rm ref}$ is denoted by
$\Delta U$, as in the free-energy approach. The 
change of any quantity resulting from the
replacement of $U_{\rm ref}$ by $U_{\rm AI}$
will also be indicated
by $\Delta$; for example, the shift of melting temperature
at a given pressure is called $\Delta T_{\rm m}$. The
melting temperature is shifted because the liquid and
solid Gibbs free energies $G^l ( p , T )$ and $G^s ( p , T )$,
and hence their difference $G^{l s} ( p , T ) \equiv
G^l ( p , T ) - G^s ( p , T )$, are shifted.  Working at the
given pressure, we take the variable $p$ as read and express
the {\em ab initio} value of $G^{l s}$ as:
\begin{equation}
G_{\rm AI}^{l s} ( T ) = G_{\rm ref}^{l s} +
\zeta \Delta G^{l s} ( T ) \; ,
\end{equation}
where the parameter $\zeta$ is introduced so that the reference
melting temperature $T_{\rm m}^{\rm ref}$ can be
written as a power series:
\begin{equation}
T_{\rm m}^{\rm AI} = T_{\rm m}^{\rm ref} + \zeta T_{\rm m}^\prime +
\zeta^2 T_{\rm m}^{\prime \prime} + \ldots \; .
\end{equation}
Since the Gibbs free energies are equal in the two phases, this
$T_{\rm m}^{\rm AI}$ is the solution of 
$G_{\rm AI}^{l s} ( T ) = 0$, which is:
\begin{equation}\label{eqn:deltat}
G_{\rm ref}^{l s} \left( T_{\rm m}^{\rm ref} + \zeta T_{\rm m}^\prime +
\zeta^2 T_{\rm m}^{\prime \prime} \ldots \right) +
\zeta \Delta G^{l s} \left( T_{\rm m}^{\rm ref} +
\zeta T_{\rm m}^\prime + \zeta^2 T_{\rm m}^{\prime \prime} + \ldots
\right) = 0 \; .
\end{equation}
Expanding in powers of $\zeta$ and equating powers, one obtains
formulas for $T_{\rm m}^\prime$, $T_{\rm m}^{\prime \prime}$, etc:
\begin{eqnarray}
T_{\rm m}^\prime & = & \Delta G^{l s} \left( T_{\rm m}^{\rm ref} \right) 
\left/ S_{\rm ref}^{l s} \right. \nonumber \\
T_{\rm m}^{\prime \prime} / T_{\rm m}^\prime & = & 
- \frac{1}{S_{\rm ref}^{ls}} \left[ T_{\rm m}^\prime C_{p , {\rm ref}}^{ls} /
2 T_{\rm m}^{\rm ref} + \Delta S^{ls} \right]  \; ,
\label{eqn:tm_shift}
\end{eqnarray}
where $S_{\rm ref}^{l s} \equiv S_{\rm ref}^l - S_{\rm ref}^s$ is
the reference entropy of fusion, $C_{p , {\rm ref}}^{ls} \equiv
C_{p , {\rm ref}}^l - C_{p , {\rm ref}}^s$ is the liquid-solid
difference of the constant-pressure specific heats, and
$\Delta S^{ls}$ is the shift of the entropy of fusion. 
We note particularly the implication of the
formula for $T_{\rm m}^\prime$. Since entropies of fusion are
on the order of $k_{\rm B}$ per atom, then a difference
$\Delta G^{l s}$ of 10~meV/atom implies a shift of melting
temperature of {\em ca.}~100~K, so that substantial errors will
need to be corrected for unless the reference total energy function
matches the {\em ab initio} one very precisely. Although we have included
a formula for $T_{\rm m}^{\prime \prime}$, it seems unlikely
that this will be used in practice, except perhaps for a rough
estimate of $T_{\rm m}^{\prime \prime} / T_{\rm m}^\prime$,
since $\Delta S^{ls}$ would be difficult to compute
without rather extensive free-energy calculations.

If simulations are done in the isothermal-isobaric ensemble,
then for closely matching $U_{\rm AI}$ and $U_{\rm ref}$, with
small fluctuations of $\Delta U$, the Gibbs free energy shifts
$\Delta G^l$ and $\Delta G^s$ can be evaluated using the well-known
expansion:
\begin{equation}
\Delta G = \langle \Delta U \rangle_{\rm ref} - 
\frac{1}{2} \beta \langle \delta \Delta U^2 \rangle_{\rm ref} + \ldots \; ,
\label{eqn:zwanzig}
\end{equation}
where $\delta \Delta U \equiv \Delta U - \langle \Delta U \rangle_{\rm ref}$,
and the averages are taken in the reference ensemble.
The {\em ab initio} simulations presented later were performed
in the isothermal-isochoric ensemble, so that the
quantity that is readily evaluated is $\Delta F ( V , T )$,
the change of Helmholtz free energy when $U_{\rm ref}$ is
replaced by $U_{\rm AI}$ are constant volume $V$. This
$\Delta F$ is given by the same formula 
Eqn~(\ref{eqn:zwanzig}), but with the
averages evaluated in the isothermal-isochoric ensemble. In this case,
we need to consider the relation between $\Delta G$ and $\Delta F$,
which is readily shown to be:
\begin{equation}
\Delta G = \Delta F - V \kappa_T \Delta p^2 \; ,
\label{eqn:deltag}
\end{equation}
where $\kappa_T$ is the isothermal compressibility and
$\Delta p$ is the change of pressure when $U_{\rm ref}$ is replaced
by $U_{\rm AI}$ at constant $V$ and $T$.

\subsection{Size effects}
\label{sec:size}

Whichever approach is used, coexistence must be treated in the
thermodynamic limit. In the free-energy approach,
almost all the free energy is that of the reference system, for which
size errors can be made negligible by doing simulations on very large
systems. Appreciable size errors remain only in the difference $F_{\rm
AI} - F_{\rm ref}$ between the {\em ab initio} and reference
systems. These can only be assessed and corrected for by explicitly
calculating $\Delta F \equiv \Delta F \equiv F_{\rm AI} - F_{\rm ref}$ 
for systems of increasing size,
as we report elsewhere for the cases of 
Al~\cite{vocadlo02} and Fe~\cite{alfe02}. Similarly, in the
coexistence approach, explicit coexistence simulations are performed
with the model system on very large systems, so that size errors are
made negligible. However, the shift of melting temperature due to the
difference $U_{\rm AI} - U_{\rm ref}$ requires calculations of
$\Delta F$ of exactly the same kind as are needed
in the free-energy approach. This means that the size errors are
essentially the same in the two approaches. We give a quantitative
assessment of size errors in the melting properties of Fe by the
coexistence approach in the next Section.

\section{The melting properties of iron}
\label{sec:iron}

\subsection{Technical details}
\label{sec:details}

Our coexistence calculations on high-pressure Fe 
use precisely the same DFT techniques
used in our free-energy work~\cite{alfe02,alfe01}, 
so we give only a brief summary
here. The exchange-correlation functional $E_{\rm xc}$
is the generalized gradient approximation known as 
Perdew-Wang 1991~\cite{wang91,perdew92}.
We use the projector-augmented-wave (PAW) implementation of 
DFT~\cite{blochl94,kresse99,alfe99b},
a technique that shares the properties both of all-electron
methods such as full-potential linearized augmented plane waves
(FLAPW)~\cite{wei85} and the ultrasoft pseudopotential 
method~\cite{vanderbilt90}. The calculations were done using the 
VASP code~\cite{kresse96a,kresse96b}.
Details of the core radii, augmentation-charge cut-offs, etc. are
exactly as in our PAW work on liquid Fe~\cite{alfe99b}.  Our division
into valence and core states is also the same: the 3$p$ electrons are
treated as core states, but their response to the high compression is
represented by an effective pair potential, with the latter
constructed using PAW calculations in which the 3$p$ states are
explicitly included as valence states. 

The reference model for our coexistence simulations
is the embedded-atom model (EAM)
recently used by Belonoshko {\em et al.}~\cite{belonoshko00} to
calculate the high-pressure melting curve of Fe. This EAM has the
standard form~\cite{daw93}, in which the total energy $E_{\rm tot} =
\sum_i E_i$ is a sum of energies $E_i$ of atoms $i$, with 
each $E_i = E_i^{\rm rep} + F ( \rho_i )$
consisting of two parts: first, a purely repulsive energy $E_i^{\rm
rep}$ represented as a sum of inverse-power pair potentials $E_i^{\rm
rep} = \sum_j^\prime \epsilon ( a / r_{ij} )^n$, where $r_{ij}$ is the
interatomic separation and the sum excludes $i=j$; second, an
`embedding' part $F ( \rho_i )$ which accounts for the metallic
bonding mainly due to partial filling of the $d$-bands. The embedding
function $F ( \rho )$ is represented as $- \epsilon C \rho^{1/2}$, and
the density $\rho_i$ for atom $i$ is given by the sum over neighbours
$\rho_i = \sum_j^\prime ( a / r_{ij} )^m$.  The parameters in this EAM
were determined in Ref.~\cite{belonoshko00} by fitting to
full-potential linearized muffin-tin orbital (FPLMTO) energies for
typical configurations of liquid iron, and they are: $n=8.137,
m=4.788, \epsilon=0.0173$~eV, $a=3.4714$~\AA, and $C=24.939$.

The aim of our calculations is not to generate the entire
melting curve, but to obtain a point on this curve at a pressure
close to the value of 330~GPa at the boundary between the Earth's solid inner
core and liquid outer core~\cite{poirier91}.
Our main simulations have been done on cells containing 8000 atoms,
constructed as a $10 \times 10 \times 40$ h.c.p. supercell. The
long axis is perpendicular to the basal plane of the crystal,
and we refer to it as the $z$-axis.
We have tested the adequacy of this system size by performing
coexistence calculations on systems of up to 20,480 atoms. 
Within the statistical errors, we were unable to detect any
difference between results for coexistence pressure and temperature
with this system size and those with the 8,000-atom system. The
mean volume per atom for the calculations reported 
here was $v = 7.12$~\AA$^3$, which is near the volumes of the
solid and liquid in the pressure region of interest. 
Since our calculations are performed at constant volume
and constant cell shape (see Sec.~\ref{sec:coexist}), we 
checked that non-hydrostatic stresses do
not affect the melting properties of the system. To do this, we performed
the simulations with the two $c / a$ values 1.64 and 1.66,
which give the $P_{zz}$
diagonal component of the stress tensor slightly larger or smaller than
the $P_{xx}$ and $P_{yy}$ components in the two cases. We find that the
effect on the melting temperature is undetectably small.

In preparing the coexisting solid and liquid as described in
Sec.~\ref{sec:coexist}, the system was initially equilibrated at
6100~K, and the high temperature used to ensure complete melting of
half the system was $5\times 10^4$~K. Once melting had been achieved,
the liquid part was re-equilibrated at 6100~K before free evolution of
the whole system was started.

\subsection{Results}
\label{sec:results}

\begin{figure}
\psfig{figure=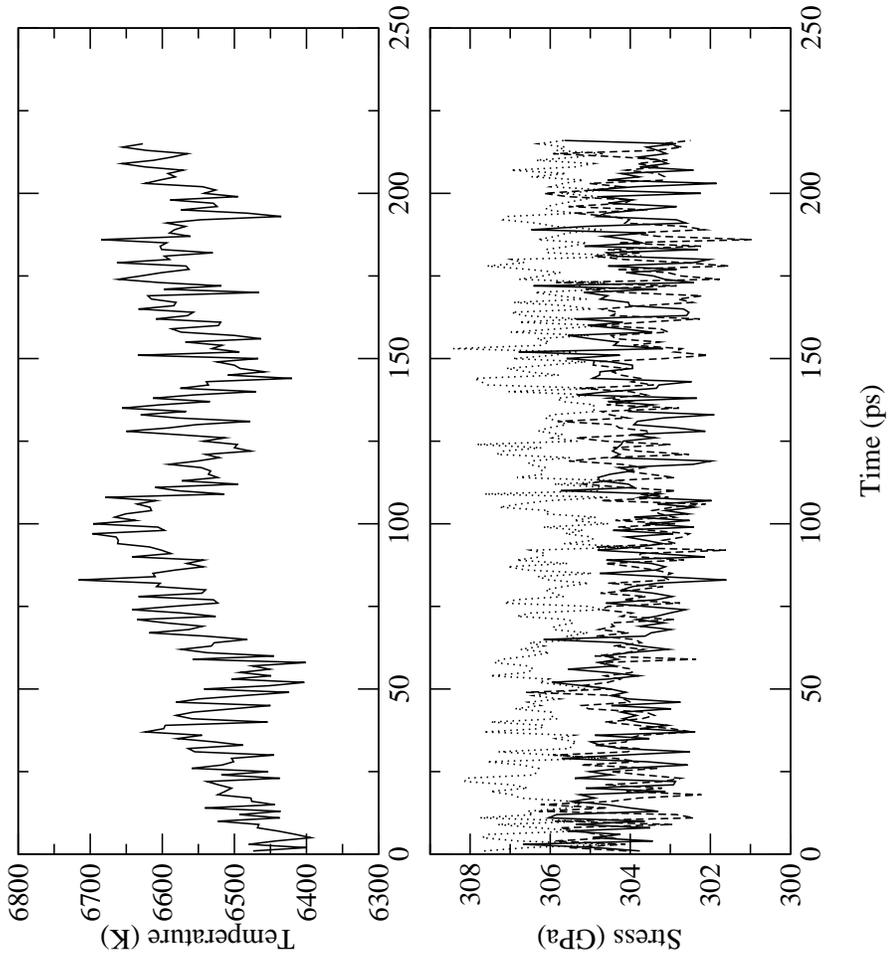,height=7in}
\caption{Time variation of temperature (upper panel) and the
three components of stress tensor $P_{xx}$ (solid curve),
$P_{yy}$ (dashed curve) and $P_{zz}$ (dotted curve) (upper panel), during 
a simulation of solid and liquid Fe coexisting at a pressure
of 305~GPa. Simulations were performed on a system of 8000 atoms
using the embedded-atom potential of Belonoshko 
{\em et al.}~\cite{belonoshko00}, with $c/a$ ratio of the h.c.p.
solid equal to 1.64.
}
\label{fig:T}
\end{figure}

\begin{figure}
\psfig{figure=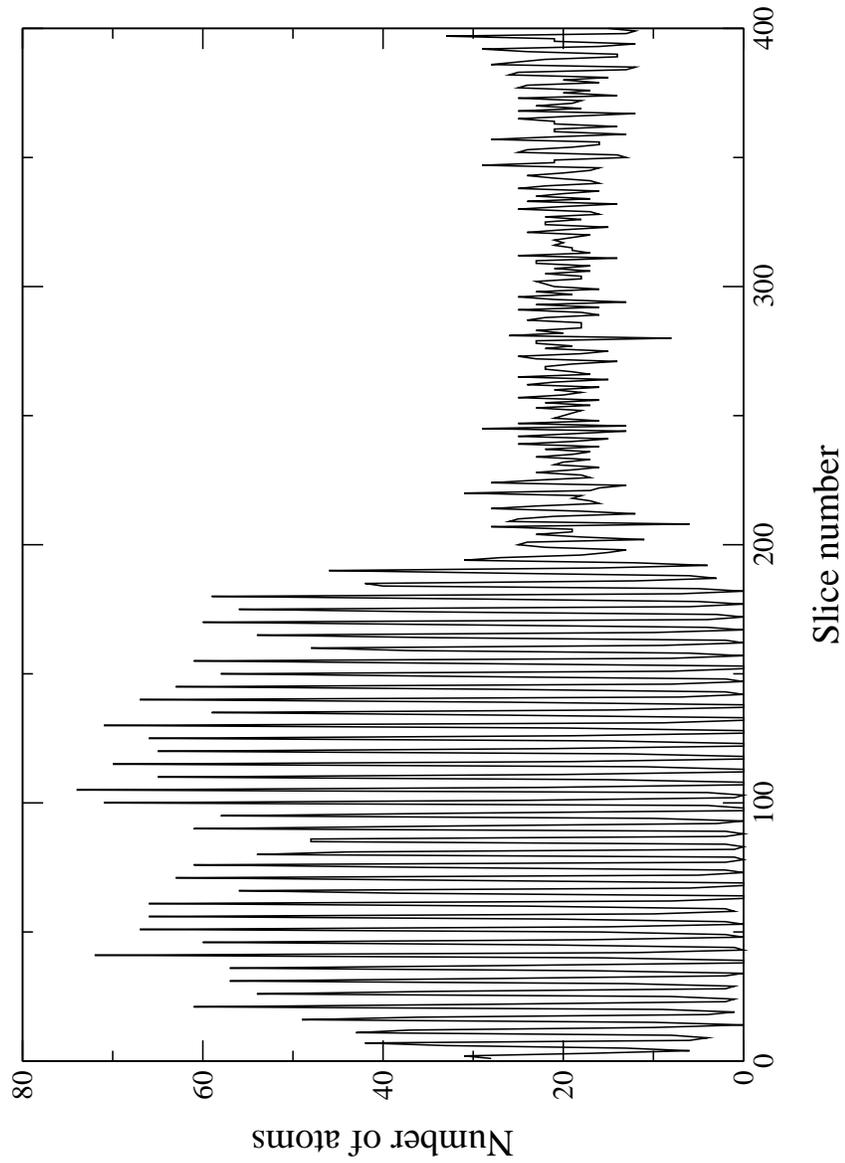,height=7in}
\caption{Density profile in a simulation of solid and liquid Fe
coexisting at a pressure of 305~GPa. The system is divided
into slices of equal thickness (0.35~\AA) parallel to the solid-liquid
interface, and graph shows number of atoms in each slice.
Simulations were performed on a system of 8000 atoms using
the embedded-atom potential of Belonoshko {\em et al.}~\cite{belonoshko00}.
}
\label{fig:density}
\end{figure}

In Fig.~\ref{fig:T}, we show the temperature of the system as a
function of simulation time for $c/a = 1.64$. We also show
the three diagonal components of the stress tensor; the off-diagonal
components fluctuate around their average value of zero, so there is
no shear stress on the cell.  After an initial equilibration period,
one sees that the temperature and the pressure settle around the
values $T=6550 \pm 100$~K and $p= 305 \pm 1$~GPa. In 
Fig.~\ref{fig:density}, we display the density profile, calculated by
dividing the simulation cell into 400 slices parallel to the
liquid-solid interface and counting the number of atoms present in
each slice. The profile shown corresponds to the last configuration in the
simulation with $c/a=1.64$; a similar profile is observed in the
simulation with $c/a=1.66$. It is easy to identify the solid and the
liquid regions in the system: in the solid region the density is a
periodic function of slice number, but in the liquid it fluctuates
randomly around its average value.  This confirms that we do indeed
have coexisting solid and liquid in the cell. At 305~GPa, the melting
temperature reported by Belonoshko with precisely the same EAM is
6680~K, which agrees with our value within the combined statistical
errors.  



We now correct the EAM reference melting 
temperature to obtain the fully {\em ab
initio} melting temperature using the techniques presented in
Sec.~\ref{sec:corrections}. In order to assess possible system-size
errors, the free-energy corrections of 
Eqs.~(\ref{eqn:zwanzig},\ref{eqn:deltag}) were
calculated with systems of 64, 150 and 288 atoms.  To do this, we used
the EAM reference model to generate long simulations for the solid and
liquid separately. From these simulations we extract typically 50 and
100 statistically independent configurations for solid and liquid
respectively, for which we calculate the DFT total energies $U_{\rm
AI}$. The differences $\Delta U \equiv U_{\rm AI} - U_{\rm ref}$ are
then used to compute the free-energy corrections and hence the shift
of melting temperature (see Eq.~(\ref{eqn:tm_shift})).
The temperature was set equal to the value of
6550~K that emerges from the coexistence simulations, and volumes per
atom of $v_s = 7.05$ and $v_l = 7.218$~\AA$^3$/atom were also deduced
from the coexistence simulations; the pressure in both cases
was 303~GPa.  (The very small difference from the pressure
of 305~GPa mentioned above is of no consequence.)
From these two volumes we can extract the volume change
on melting, which is $\Delta V \sim 2.5\%$, the same value
reported as by Belonoshko {\em et al.}~\cite{belonoshko00}. 
(The value from our {\em ab initio} free-energy
work was 1.8~\% at 305~GPa~\cite{alfe02}.) We also
calculated the entropy of melting from the relation $T S_{\rm ref}^{ls} =
E_{\rm ref}^{ls} + p V_{\rm ref}^{ls}$; the energy change on melting
$E_{\rm ref}^{ls}$ was calculated using the two separate simulations of
solid and liquid at $p=303$ GPa and $T=6550 $K, and we obtained
$S_{\rm ref}^{ls} = 0.88 $~k$_B$/atom; a similar value of $S^{ls}$ can be
deduced from the work of Belonoshko {\em et al.}~\cite{belonoshko00} by
using their value for $V^{ls}$ and the Clausius-Clapeyron relation
on the reported melting curve. (The value from our {\em ab initio}
free-energy work was 1.07~$k_{\rm B}$/atom at 305~GPa~\cite{alfe02}.)
The DFT energies were carefully
checked for electronic $k$-point errors, as in our free-energy
work~\cite{alfe02,alfe01}. As expected 
from that work, $\Gamma$-point sampling for the
64-atom system underestimates the fully converged energies of liquid and
solid by {\em ca.}~10 and 50~meV/atom respectively. With 150 atoms,
$\Gamma$-point sampling gives negligible errors for the liquid, but an
error of {\em ca.}~$- 8$~meV/atom for the solid. With 288 atoms, we
have used only $\Gamma$-point sampling, but the indications are that
$k$-point errors should be negligible for both phases.

We report in Table~\ref{tab:results} the results for 
$\langle \Delta U \rangle_{\rm ref}$ and 
$\langle (\delta\Delta U)^2 \rangle_{\rm ref}$ for the
three system sizes, the results being already corrected
for electronic $k$-point errors. An important feature of these results is
that there is no discernible system size effect on these corrections
within the statistical errors of 5~meV or less, so that
systems of 64 atoms are adequate for calculating the
corrections. From Eq.~(\ref{eqn:tm_shift}), this imples
that the shift of $T_{\rm m}$ due to size errors will not more
than $\sim 50$~K.
In order to obtain the corrections $\Delta G$ to the Gibbs free
energy, we need to include the term in $\Delta p^2$ in
Eq.~(\ref{eqn:deltag}). We find that the pressure differences between
the EAM and {\em ab-initio} systems are only 22 and 12 GPa for solid
and liquid respectively, which give this term values of {\em
ca.}~5~meV.  From Table~\ref{tab:results}, we see that the free-energy
differences between the {\em ab initio} and EAM systems have the
effect of stabilizing the liquid with respect to the solid by about
35~meV/atom compared with the EAM.
In order to obtain the correction to the melting temperature from
Eq.~(\ref{eqn:tm_shift}), we use the value for the entropy of
melting $S_{\rm ref}^{ls}$ of $0.88 k_{\rm B}$/atom quoted above.
This gives the first-order correction $T_{\rm m}^\prime = - 450$~K,
so that our corrected $T_m$ at 305~GPa is $6100 \pm 100$~K. This result
is in very close agreement with the free-energy approach, which at
$p=305$~GPa gives $T_m = 6100$~K. (This value is somewhat lower
than the preliminary result of our free-energy 
calculations~\cite{alfe99}; the
downward revision came from a careful reanalysis of
the anharmonic free energy of the h.c.p. crystal, as reported
in Refs.~\cite{alfe02,alfe01}.)

Although the EAM of Belonoshko {\em et al.}~\cite{belonoshko00} mimics
our {\em ab initio} systems reasonably well, we have found that the
model can be still further improved by refitting it to our {\em ab
initio} energies of the solid and liquid. In doing this, we allowed
only the repulsive potential and the strength of the embedding energy
to change, so that only the parameters $n$, $\epsilon$ and $C$ are
allowed to vary. These were adjusted to minimise the $\delta \Delta U$
fluctuations for the solid, while also maintaining the correct
pressure.
The new parameters are $n=5.93, \epsilon=0.1662$~eV, $C=16.55$, the
parameters $a=3.4714$~\AA\ and $m=4.788$ retaining their original
values.  We then repeated the coexistence simulation using this new
EAM and obtained a coexistence temperature of $T_m = 6200 \pm
100$~K at a pressure of $p = 323.5 \pm 1$ GPa. The very small
free-energy corrections (Table~\ref{tab:results2}) slightly stabilize
the solid, resulting in an increase of melting temperature of {\em
ca.}~50~K to give $6250 \pm 100$~K. The differences with the {\em
ab-initio} pressures are negligible in this case, and the term in
$\Delta p^2$ in Eq.~(\ref{eqn:deltag}) does not contribute to the free
energy difference.  At this pressure, the melting temperature from our
free energy approach was $T_m = 6290$~K~\cite{alfe02}, 
so that once again we find
very good agreement between the two approaches.

\section{Discussion and conclusions}
\label{sec:discuss}

We have advocated the aim of calculating the melting properties that
follow from a chosen approximation for the exchange-correlation
functional $E_{\rm xc}$, with the errors due to all other
approximations being made negligible. We have stressed that in the
free-energy approach the total energy of the reference model does not
have to agree exactly with the {\em ab initio} total energy, since
this approach includes the calculation of the free energy difference
between the two systems. Up to now, no allowance has been made for
such differences in the coexistence approach, but we have shown how
corrections can be made so that the final results are again
independent of the reference model. Our practical results for
high-pressure Fe demonstrate that the two approaches then give the
same melting temperatures, as would be expected.

Our analysis has implications for the way in which reference systems
are constructed. In the free-energy approach, the crucial requirement
of a reference model is that the fluctuations of the difference
$\Delta U$ between the {\em ab initio} and reference energies be as
small as possible; we have the freedom to use different reference
models for the solid and liquid phases, and a constant offset between
the {\em ab initio} and reference energies is of no consequence. In
the coexistence approach, it seems clear that the model should be
chosen so that the $\Delta G$ corrections, and the resulting shift of
melting temperature away from that of the reference system, be as
small as possible. This demands more of the reference system than in
the free-energy approach. It is necessary but not sufficient that the
$\Delta U$ fluctuations be small, in order that $\Delta G$ be small
(see Eq.~(\ref{eqn:deltag})). In addition, a constant energy offset
will shift $T_{\rm m}$, if the offset is not identical in the two
phases, since the same model must simultaneously reproduce the
energetics of both. This implies that 
force matching~\cite{laio00,ercolessi94} may not always be
a reliable way of constructing reference models: even if the reference
and {\em ab initio} forces match precisely in both phases, there may
still be a non-cancelling $\Delta U$ offset that will shift $T_{\rm
m}$.  In this sense, energy matching is safer than force matching.

Concerning finite-size errors, we have shown that these affect
appreciably only the free-energy differences between the
{\em ab initio} and reference systems, since in both approaches
the calculations on the reference model will always be done on
systems large enough to make size errors negligible. If anything,
these errors may be more troublesome in the coexistence
approach, because of the need to make the reference model fit
both phases at once, so that the free-energy corrections may
sometimes be larger. However, in our coexistence calculations
on high-pressure Fe, we have shown that these errors are
unlikely to shift $T_{\rm m}$ by more than $\sim$~50~K.

In commenting on the strengths and weaknesses of the two approaches,
we note that they differ mainly in the way of treating the
thermodynamic properties of the reference model. In both approaches,
corrections are then needed to obtain the properties of the {\em ab
initio} system, and these involve {\em ab initio} simulations to
perform thermodynamic integration or to compute the free-energy
differences of Eq.~(\ref{eqn:zwanzig}). The melting properties of the
reference model are probably more straightforward to calculate by the
coexistence approach, since the free-energy approach requires rather
intricate thermodynamic integrations.  However, the heaviest
computational effort comes in the calculation of the free-energy
corrections, and here we believe the free-energy approach may have the
advantage, since the effort can be reduced by using different
reference systems for the two phases. In practice, we have found it
very helpful to use both approaches, as in the work on Fe reported
here.

\section*{Acknowledgments}

The work of DA is supported by a Royal Society University
Research Fellowship, and the work of MJG by Daresbury Laboratory
and GEC. The computational facilities of the UCL HiPerSPACE Centre (JREI
grant JR98UCGI) and the Manchester CSAR Service (grant GST/02/1002 to
the Minerals Physics Consortium) were used.
The authors thank A.~Laio and S.~Scandolo for
detailed discussions about their coexistence calculations.

\pagebreak

\begin{table}
\begin{tabular}{l|cccc}
& \multicolumn{2}{c}{$\langle \Delta U \rangle_{\rm ref} / N$} & 
\multicolumn{2}{c}{$\langle (\delta \Delta U)^2 \rangle_{\rm ref} / N$} \\
$N$ & Liquid & Solid & Liquid & Solid \\
\hline
64 & $-6.940 \pm 0.003$ & $ -6.909 \pm 0.002 $ & $ 0.023 \pm 0.004$ & $ 0.014 \pm 0.003$ \\
150 & $-6.934 \pm 0.001$ & $ -6.912 \pm 0.001 $ & $ 0.023 \pm 0.003$ & $ 0.009 \pm 0.002$ \\
288 & $-6.939 \pm 0.001$ & $ -6.909 \pm 0.001 $ & $ 0.024 \pm 0.004$ & $ 0.014 \pm 0.003$ \\
\end{tabular}
\caption{Thermal average $\langle \Delta U \rangle_{\rm ref}$ of the
difference $\Delta U \equiv U_{\rm AI} - U_{\rm ref}$ of {\em ab initio}
and reference energies, and thermal average 
$\langle ( \delta \Delta U )^2 \rangle_{\rm ref}$ of the
squared fluctuations of $\delta \Delta U \equiv \Delta U - 
\langle \Delta U \rangle_{\rm ref}$, with averages evaluated in
the ensemble of the reference system and normalized by dividing
by the number of atoms $N$ (eV units). Results from simulations with
difference $N$ are reported for liquid and solid Fe for the
thermodynamic state $p = 303$~GPa, $T = 6550$~K. Reference
model is the embedded-atom model of 
Belonoshko {\em et al.}~\cite{belonoshko00}.
}
\label{tab:results}
\end{table}

\begin{table}
\begin{tabular}{l|cccc}
& \multicolumn{2}{c}{$\langle \Delta U \rangle_{\rm ref} / N$} & 
\multicolumn{2}{c}{$\langle (\delta \Delta U)^2 \rangle_{\rm ref} / N$} \\
$N$ & Liquid & Solid & Liquid & Solid \\
\hline
64 & $7.213 \pm 0.002$ & $ 7.200 \pm 0.002 $ & $ 0.018 \pm 0.003$ & $ 0.009\pm 0.002$ \\
\end{tabular}
\caption{Thermal averages $\langle \Delta U \rangle_{\rm ref}$ and
$\langle ( \delta \Delta U )^2 \rangle_{\rm ref}$ (eV units, see caption of
Table~1) for the EAM reference model obtained by refitting that of
Belonoshko {\em et al.}~\cite{belonoshko00}.
}
\label{tab:results2}
\end{table}

\end{document}